\documentclass[aps,prd,twocolumn,superscriptaddress,nofootinbib,showpacs,showkeys]{revtex4-1}
\usepackage[colorlinks,citecolor=blue]{hyperref}
\usepackage{amsmath}
\usepackage{amssymb}
\usepackage{graphicx}
\usepackage{slashed}
\usepackage{xcolor}
\usepackage{array}
\usepackage{multirow}
\usepackage{cancel}
\usepackage{color}
\usepackage{mathtools}
\usepackage{listings}
\usepackage{xcolor}
\usepackage{float}
\usepackage{slashed}

\usepackage{amsmath}%\ge=\geqslant; \le=\leqslant
\usepackage{wrapfig}

\usepackage{tikz}
\usetikzlibrary{arrows,shapes}
\usetikzlibrary{trees}
\usetikzlibrary{matrix,arrows} 				% For commutative diagram
\usetikzlibrary{positioning}				% For "above of=" commands
\usetikzlibrary{calc,through}				% For coordinates
\usetikzlibrary{decorations.pathreplacing}  % For curly braces
\usepackage{pgffor}							% For repeating patterns

\usetikzlibrary{decorations.pathmorphing}	% For Feynman Diagrams
\usetikzlibrary{decorations.markings}
\tikzset{
    vector/.style={decorate, decoration={snake}, draw},
	provector/.style={decorate, decoration={snake,amplitude=2.5pt}, draw},
	antivector/.style={decorate, decoration={snake,amplitude=-2.5pt}, draw},
    fermion/.style={draw=black, postaction={decorate},
        decoration={markings,mark=at position .55 with {\arrow[draw=black]{>}}}},
    fermionbar/.style={draw=black, postaction={decorate},
        decoration={markings,mark=at position .55 with {\arrow[draw=black]{<}}}},
    fermionnoarrow/.style={draw=black},
    gluon/.style={decorate, draw=black,
        decoration={coil,amplitude=4pt, segment length=5pt}},
    scalar/.style={dashed,draw=black, postaction={decorate},
        decoration={markings,mark=at position .55 with {\arrow[draw=black]{>}}}},
    scalarbar/.style={dashed,draw=black, postaction={decorate},
        decoration={markings,mark=at position .55 with {\arrow[draw=black]{<}}}},
    scalarnoarrow/.style={dashed,draw=black},
    electron/.style={draw=black, postaction={decorate},
        decoration={markings,mark=at position .55 with {\arrow[draw=black]{>}}}},
	bigvector/.style={decorate, decoration={snake,amplitude=4pt}, draw},
}

\tikzstyle{block} = [draw, rectangle, 
    minimum height=3em, minimum width=6em]

\usepackage{hyperref}			% Has to be at the end.

\newcommand{\be}{\begin{equation}}
\newcommand{\ee}{\end{equation}}
\newcommand{\beq}{\begin{equation}}
\newcommand{\eeq}{\end{equation}}
\newcommand{\bea}{\begin{eqnarray}}
\newcommand{\eea}{\end{eqnarray}}
\newcommand{\besp}{\begin{equation}\begin{split}}
\newcommand{\eesp}{\end{split}\end{equation}}

\newcommand{\nn}{\nonumber}

\newcommand{\Eq}[1]{Eq.~(\ref{#1})}
\newcommand{\Dfbd}{\mathord{\buildrel{\lower3pt\hbox{$\scriptscriptstyle\leftrightarrow$}}\over {D}_{\mu}}}
\newcommand{\ave}[1]{\left\langle #1\right\rangle}
\newcommand{\abs}[1]{\left| #1\right|}

\def\hc{{\rm h.c.}}
\def\mL{\mathcal{L}}

\def\mO{\mathcal{O}}

\begin{document}
\title{Testing leptogenesis at the LHC and future muon colliders: a $Z'$ scenario}

\author{Wei Liu}
\email{wei.liu@njust.edu.cn}
\affiliation{Department of Applied Physics, Nanjing University of Science and Technology, Nanjing 210094, People's Republic of China}

\author{Ke-Pan Xie}
\email{kepan.xie@unl.edu}
\affiliation{Department of Physics and Astronomy, University of Nebraska, Lincoln, NE 68588, USA}

\author{Zihan Yi}
\email{zhygw@njust.edu.cn}
\affiliation{Department of Applied Physics, Nanjing University of Science and Technology, Nanjing 210094, People's Republic of China}

\begin{abstract}

If the masses of at least two generations of right-handed neutrinos (RHNs) are near-degenerate, the scale of leptogenesis can be as low as $\sim$ 100 GeV. In this work, we study probing such resonant leptogenesis in the $B-L$ model at the LHC and future multi-TeV muon colliders via the process $Z'\to NN\to\ell^\pm\ell^\pm+{\rm jets}$, with $Z'$ the $U(1)_{B-L}$ gauge boson and $N$ the RHN. The same-sign dilepton feature of the signal makes it almost background-free, while the event number difference between positive and negative leptons is a hint for $CP$ violation, which is a key ingredient of leptogenesis. We found that resonant leptogenesis can be tested at the HL-LHC for $M_{Z'}$ up to 12 TeV, while at a 10 (30) TeV muon collider the reach can be up to $M_{Z'}\sim30~(100)$ TeV via the off-shell production of $Z'$.

\end{abstract}

%%%%%%%%%%%%%%%%%%%%%%%%%%%%%%%%%%%%%%%%%%%%%%%%%%%%%%%%%%%%%%%%%%%%%%%%%%%%%%%%%
\maketitle
\section{Introduction}

The baryon asymmetry of the Universe (BAU) is one of the most mysterious unsolved problems in the Standard Model (SM) of particle physics. Leptogenesis is a very attractive explanation~\cite{Fukugita:1986hr,Luty:1992un,Davidson:2008bu}, as it links the BAU to the origin of neutrino masses. In that mechanism, the $CP$ violating decay of the heavy right-handed neutrinos (RHN) to the SM leptons generates a lepton asymmetry, which is then converted to the BAU via the electroweak (EW) sphaleron process. The same interaction also accounts for the tiny neutrino masses~\cite{Davis:1994jw, Fukuda:1998mi,Eguchi:2002dm} via the Type-I seesaw mechanism~\cite{Minkowski:1977sc}. In the conventional thermal leptogenesis formalism, the $CP$ violating effects are related to the RHN mass $M_N$, and the explanation of the BAU requires $M_N\gtrsim10^9$ GeV~\cite{Davidson:2002qv}, making it very challenging to test the mechanism experimentally. However, the constraints on $M_N$ can be relaxed if at least two of the RHNs are highly degenerate, so that the $CP$ asymmetry is resonantly enhanced, and hence even $\mO({\rm TeV})$ RHNs can generate the observed BAU~\cite{Flanz:1996fb,Pilaftsis:1997jf,Pilaftsis:2003gt,Iso:2010mv,Dev:2017wwc}. In particular, if such a resonant leptogenesis mechanism is embedded into a gauge theory in which the RHNs exist naturally for anomaly cancellation, then the $\mO({\rm TeV})$ leptogenesis is testable at the colliders via searches for RHNs or new gauge/scalar bosons as well as the leptonic charge asymmetries~\cite{Blanchet:2009bu,Okada:2012fs,Heeck:2016oda,Dev:2017xry, BhupalDev:2014hro, Chun:2017spz, BhupalDev:2015khe, Dev:2019rxh}.\footnote{Also at Ref.~\cite{Chauhan:2021xus} for a recent review.}

In this article, we perform a comprehensive study of the collider phenomenology of the resonant leptogenesis mechanism in the gauged $U(1)_{B-L}$ extended SM (i.e. the so-called $B-L$ model~\cite{Davidson:1978pm,Marshak:1979fm,Mohapatra:1980qe,Davidson:1987mh}), focusing on the interplay between leptogenesis and the $Z'$ gauge boson of the $U(1)_{B-L}$ group, since the $Z'$-mediated scattering process greatly impacts the generated BAU. In particular, we use the $Z'\to NN\to\ell^\pm\ell^\pm+{\rm jets}$ channel to probe the $Z'$, $N$ particles as well as the $CP$ violation. Compared with previous studies on similar topics~\cite{Blanchet:2009bu,Okada:2012fs,Heeck:2016oda,Dev:2017xry, BhupalDev:2014hro, Chun:2017spz, BhupalDev:2015khe, Dev:2019rxh}, our work includes not only the newest constraints from the LHC and the corresponding projected reach at the future HL-LHC, but also the first projections at the future multi-TeV muon colliders. The study on the physics potential of muon colliders started around three decades ago~\cite{Barger:1995hr,Barger:1996jm}, and receives a renewed interest recently~\cite{Han:2012rb,Chakrabarty:2014pja,Ruhdorfer:2019utl,DiLuzio:2018jwd,Delahaye:2019omf,Long:2020wfp,Buttazzo:2018qqp,Costantini:2020stv,Han:2020uid,Capdevilla:2020qel,Han:2020uak,Han:2020pif,Bartosik:2020xwr,Chiesa:2020awd,Yin:2020afe,Buttazzo:2020eyl,Lu:2020dkx,Huang:2021nkl,Liu:2021jyc,Cheung:2021iev,Han:2021udl,Capdevilla:2021fmj,Huang:2021biu,Han:2021kes,AlAli:2021let,Asadi:2021gah,Franceschini:2021aqd,Buarque:2021dji,Han:2021lnp,Chiesa:2021qpr,Bandyopadhyay:2021pld,Sen:2021fha}. Due to the high energy and precision measurement environment, the future multi-TeV muon colliders offer us the opportunity to probe both SM and beyond the SM physics very accurately. 

This paper is organized as follows. The $B-L$ model and the corresponding (resonant) leptogenesis is described in Section~\ref{sec:blz}, where the relation between BAU and the size of $CP$ asymmetry is derived. Then we study the collider phenomenology in Section~\ref{sec:mc}, including the reach for $Z'$ boson and the $CP$ asymmetry (via RHNs) at the HL-LHC and muon colliders. Finally, the conclusion is given in Section~\ref{sec:con}.

\section{Resonant leptogenesis in the $B-L$ model}\label{sec:blz}

\subsection{The model}

In the $B-L$ model, three generations of RHNs (with $B-L=-1$) are needed naturally for gauge anomaly cancellation. Besides, the model also contains an extra gauge boson $Z'$ and a complex scalar $\Phi=(\phi+i\eta)/\sqrt2$ with $B-L=2$ that breaks the $U(1)_{B-L}$ spontaneously. The relevant Lagrangian reads
\bea\label{LB-L}
\mL_{B-L}&=&\sum_{i}\bar\nu_R^ii\slashed{D}\nu_R^i-\frac12\sum_{i,j}\left(\lambda_N^{ij}\bar\nu_R^{i,c}\Phi\nu_R^j+\hc\right)\nn
\\
&&-\sum_{i,j}\left(\lambda_D^{ij}\bar\ell_L^i\tilde H\nu_R^j+\hc\right)\\
&&+D_\mu\Phi^\dagger D^\mu\Phi-\lambda_\phi\left(|\Phi|^2-\frac{v_\phi^2}{2}\right)^2-\frac14Z'_{\mu\nu}Z'^{\mu\nu},\nn
\eea
with $i$, $j$ being the family indices, $D_\mu=\partial_\mu-ig_{B-L}XZ_\mu'$ the covariant derivative ($X$ is the $B-L$ quantum number), $\ell_L$ the SM left-handed lepton doublet and $H$ the SM Higgs doublet. Without loss of generality, we assume $\lambda_N^{ij}={\rm diag}\{\lambda_{N_1},\lambda_{N_2},\lambda_{N_3}\}$, and define the four-component Majorana RHNs as $N_i=\nu_R^i+(\nu_R^i)^c$. The minimum of the scalar potential is at $\ave{|\Phi|}=v_\phi/\sqrt{2}$, breaking the $U(1)_{B-L}$ symmetry and providing masses for the particles
\be
M_{Z'}=2g_{B-L}v_\phi,~ M_{N_i}=\lambda_{N_i}\frac{v_\phi}{\sqrt2},~ M_\phi=\sqrt{2\lambda_\phi}v_\phi,
\ee
and the imaginary part $\eta$ of $\Phi$ is absorbed to be the longitudinal mode of $Z'$. We are interested in the parameter region $v_\phi\sim\mO({\rm TeV})$.

The usage of Yukawa interactions between $\ell$ and $N$ is two-fold. On one hand, they account for the tiny left-handed neutrino mass via the Type-I seesaw
\be
m_{\nu_L}\sim\frac{\lambda_D^2v^2}{M_N}\sim0.06~{\rm eV}\times\left(\frac{\lambda_D}{10^{-6}}\right)^2\left(\frac{1~{\rm TeV}}{M_N}\right),
\ee
while on the other hand they trigger the RHN decay $N\to\ell H/\bar\ell H^*$, which is the crucial process in thermal leptogenesis. Due to the $CP$ violation phase in the Yukawa couplings, the widths of $N\to\ell H$ and $N\to\bar\ell H^*$ are different, and hence a $CP$ asymmetry can be defined as~\cite{Flanz:1996fb,Pilaftsis:1997jf,Pilaftsis:2003gt,Iso:2010mv}
\bea\label{epsilon}
\epsilon_i&=&\frac{\sum_j\Gamma_{N_i\to\ell_jH}-\Gamma_{N_i\to\bar\ell_jH^*}}{\sum_j\Gamma_{N_i\to\ell_jH}+\Gamma_{N_i\to\bar\ell_jH^*}}\\
&=&-\sum_{j\neq i}\frac{M_{N_i}\Gamma_{N_j}}{M_{N_j}^2}\left(\frac{V_{ij}}{2}+S_{ij}\right)\frac{{\rm Im}(\lambda_D\lambda_D^\dagger)^2_{ij}}{(\lambda_D\lambda_D^\dagger)_{ii}(\lambda_D\lambda_D^\dagger)_{jj}},\nn
\eea
where
\be\begin{split}
V_{ij}=&~2\frac{M_{N_j}^2}{M_{N_i}^2}\left[\left(1+\frac{M_{N_j}^2}{M_{N_i}^2}\right)\ln\left(1+\frac{M_{N_j}^2}{M_{N_i}^2}\right)-1\right],\\
S_{ij}=&~\frac{M_{N_j}^2(M_{N_j}^2-M_{N_i}^2)}{(M_{N_j}^2-M_{N_i}^2)^2+M_{N_i}^2\Gamma_{N_j}^2},
\end{split}\ee
are respectively the vertex correction and RHN self-energy correction to the decay process, and the tree level width is
\be
\Gamma_{N_j}=\frac{M_{N_j}}{8\pi}(\lambda_D\lambda_D^\dagger)_{jj}.
\ee
If there is a mass hierarchy among the three RHNs, i.e. $M_{N_1}\ll M_{N_2}\ll M_{N_3}$, $V_{ij}$ and $S_{ij}$ are comparable, and $\epsilon_i$ is proportional to the lightest RHN mass but typically $\gtrsim10^{-6}$. Therefore, a sizable BAU requires a RHN with mass $\gtrsim10^9$ GeV~\cite{Davidson:2002qv}. However, if at least two RHNs are highly degenerate that $|M_{N_j}^2-M_{N_i}^2|\sim M_{N_i}\Gamma_{N_j}$, then $S_{ij}\sim M_{N_j}/\Gamma_{N_j}\gg1$, and $\epsilon_i$ can reach $\mO(1)$~\cite{Flanz:1996fb,Pilaftsis:1997jf,Pilaftsis:2003gt,Iso:2010mv,Dev:2017wwc}. In this case, even $\mO(\rm TeV)$ RHNs can generate a successful BAU. This is the scenario under consideration in this article.

\subsection{Thermal leptogenesis}

\begin{table*}\small\renewcommand\arraystretch{1.8}\centering
\begin{tabular}{|c|c|c|c|c|c|}\hline
Reaction rates & $2\to2$ scattering & $\hat\sigma(s)$ \\ \hline
$\gamma_{h,s}$ & $N_1\ell\to\bar tq$ & $\frac{3y_t^2\lambda_D^2}{4\pi}\left(\frac{x-1}{x}\right)^2$ \\ \hline
$\gamma_{h,t}$ & $N_1q\to t\ell$ or $N_1\bar t\to\bar q\ell$ & $\frac{3y_t^2\lambda_D^2}{4\pi}\left(\frac{x-1}{x}+\frac1x\ln\frac{x-1+M_h^2/M_N^2}{M_h^2/M_N^2}\right)$ \\ \hline
$\gamma_{N,s}$ & $\ell H\to\bar\ell H^*$ & $\frac{\lambda_D^4}{2\pi}\left[1+\frac{2}{D_1(x)}+\frac{x}{2D_1^2(x)}-\left(1+\frac{2(x+1)}{D_1(x)}\right)\frac{\ln(1+x)}{x}\right]$ \\ \hline
$\gamma_{N,t}$ & $\ell\ell\to H^*H^*$ & $\frac{\lambda_D^4}{2\pi}\left(\frac{x}{2(x+1)}+\frac{\ln(1+x)}{x+2}\right)$ \\ \hline
\end{tabular} 
\caption{The reaction rates and the reduced cross sections taken from Refs.~\cite{Iso:2010mv,Plumacher:1996kc}, where $x=s/M_N^2$ and $D_1(x)=x-1+(\Gamma_N^2/M_N^2)/(x-1)$.}\label{tab:reaction_rates}
\end{table*}

We assume first two flavors of RHNs are near-degenerate, i.e. $(M_{N_1}^2-M_{N_2}^2)\sim M_{N_1}\Gamma_{N_2}$, while the third flavor RHN is much heavier, i.e. $M_{N_3}\gg M_{N_1}$. The BAU receives contributions from the $CP$ violating decays of both $M_{N_1}$ and $M_{N_2}$. If the decay widths $\Gamma_{N_1}$ and $\Gamma_{N_2}$ are comparable, then the generated BAU should be twice of that from mere $N_1$ decay. However, if the decay widths have a hierarchy, e.g. $\Gamma_{N_2}\ll\Gamma_{N_1}$, then so do the $CP$ asymmetries, as $\epsilon_2\sim2\epsilon_1\Gamma_{N_2}/\Gamma_{N_1}\ll\epsilon_1$, and in the meantime the washout from $N_2$ is negligible~\cite{Iso:2010mv}. In that case, the generated BAU is dominated by $N_1$ decay, and a one-flavor discussion on $N_1$ is sufficient. For simplicity, we will consider such a scenario throughout this article. We then denote $\epsilon_1$ as $\epsilon$, and $M_{N_1}\approx M_{N_2}$ as $M_N$ from now on. In the radiation dominated era, the energy and entropy densities of the Universe are respectively
\be
\rho=\frac{\pi^2}{30}g_*T^4,\quad s=\frac{2\pi^2}{45}g_*T^3,
\ee
where $g_*=106.75$ is the number of relativistic degrees of freedom. The Hubble constant is then derived by the first Friedmann equation $H^2=(8\pi/3M_{\rm Pl}^2)\rho$, with $M_{\rm Pl}=1.22\times10^{19}$ GeV the Planck scale. Defining the dimensionless parameter $z\equiv M_N/T$, the Boltzmann equations for the RHN and net $B-L$ number yield in the thermal bath read~\cite{Iso:2010mv,Plumacher:1996kc,Perez:2021udy}
\begin{widetext}
\be\label{leptogenesis}\begin{split}
\frac{s_NH_N}{z^4}\frac{dY_N}{dz}=&~-\left(\frac{Y_N}{Y_N^{\rm eq}}-1\right)(\gamma_D+2\gamma_{h,s}+4\gamma_{h,t})-\left(\frac{Y_N^2}{(Y_N^{\rm eq})^2}-1\right)2\gamma_{Z'},\\
\frac{s_NH_N}{z^4}\frac{dY_{B-L}}{dz}=&~-\left[\frac{Y_{B-L}}{2Y_\ell^{\rm eq}}-\epsilon\left(1-\frac{Y_N}{Y_N^{\rm eq}}\right)\right]\gamma_D-\frac{Y_{B-L}}{Y_\ell^{\rm eq}}\left[2(\gamma_{N,s}+\gamma_{N,t}+\gamma_{h,t})+\frac{Y_N}{Y_N^{\rm eq}}\gamma_{h,s}\right],
\end{split}\ee
\end{widetext}
where $s_N$ and $H_N$ are the entropy density and Hubble constant at $z=1$, respectively, and the abundances are defined as number density to entropy density ratios (e.g. $Y_N=n_N/s$), with
\be
Y_N^{\rm eq}=\frac{45z^2}{2\pi^4g_*}K_2(z),\quad Y_\ell^{\rm eq}=\frac{3\zeta(3)}{2\pi^2}\frac{M_N^3}{s_N},
\ee
the equilibrium abundances, and $K_i(z)$ is the modified Bessel function of the $i$-th kind. When writing \Eq{leptogenesis}, we have neglected the charge lepton flavor effects, which can affect the production and washout of the lepton asymmetry and the low energy experiments such as $\mu\to e\gamma$. See the review~\cite{Dev:2017trv} for a fully flavor-covariant treatment on the resonant leptogenesis. In this article, since we are interested in the $\ell^\pm\ell^\pm+{\rm jets}$ final state with $\ell=e$, $\mu$, the flavor effects from the first two generation of charged leptons are expected to be subdominant.

The reaction rates in \Eq{leptogenesis} are defined as~\cite{Plumacher:1996kc},
\be
\gamma_D=\frac{M_N^3}{\pi^2}\Gamma_N\frac{K_1(z)}{z},
\ee
for the $N\to\ell H/\bar\ell H^*$ decay, and
\bea
\gamma_{ab\to cd}&\equiv&\ave{\sigma_{ab\to cd}v}n_a^{\rm eq}n_b^{\rm eq}\\
&=&\frac{M_N}{64\pi^4z}\int_{s_{\rm min}}^\infty ds\,\hat\sigma_{ab\to cd}(s)\sqrt{s}K_1\left(\frac{\sqrt{s}}{M_N}z\right),\nn
\eea
for the $2\to2$ scattering, where $s_{\rm min}=\max\{(M_a+M_b)^2,~(M_c+M_d)^2\}$, and the dimensionless reduced scattering cross section is
\begin{multline}
\hat\sigma_{ab\to cd}(s)\equiv 2\,\sigma_{ab\to cd}(s)\cdot s\cdot\\
\left[1-2\left(\frac{M_a^2}{s}+\frac{M_b^2}{s}\right)+\left(\frac{M_a^2}{s}-\frac{M_b^2}{s}\right)^2\right].
\end{multline}
The correspondence between the reaction rates in \Eq{leptogenesis} and the $2\to2$ processes are listed in Table~\ref{tab:reaction_rates}, except the case of $\gamma_{Z'}$, which corresponds to the scattering $NN\to Z'\to f\bar f$ ($f$ denotes the SM fermions) and is highlighted below
\be
\hat\sigma_{Z'}(s)=\frac{13g_{B-L}^4}{6\pi}\frac{\sqrt{x(x-4)^3}}{(x-M_{Z'}^2/M_N^2)^2+M_{Z'}^2\Gamma_{Z'}^2/M_N^4},
\ee
where
\be
\label{eq:gammaz}
\frac{\Gamma_{Z'}}{M_{Z'}}=\frac{g_{B-L}^2}{24\pi}\left[13+2\left(1-\frac{4M_N^2}{M_{Z'}^2}\right)^{3/2}\theta(M_{Z'}-2M_N)\right],
\ee
with $x=s/M_N^2$ and $\theta$ the Heaviside step function. We also assume $M_{Z'}$, $M_\phi\gtrsim M_N$ so that scatterings $NN\to Z'Z'$ or $\phi\phi$ are suppressed.\footnote{The impact of $NN\to Z'Z'$ and $NN\to\phi\phi$ can be found in Ref.~\cite{Heeck:2016oda}.}

For a $\mO(\rm TeV)$ leptogenesis, $\lambda_D\sim10^{-6}$, making the $\gamma$'s in Table~\ref{tab:reaction_rates} rather small, as they are proportional to $\lambda_D^2$ or $\lambda_D^4$. Only the reaction rate $\gamma_{Z'}$ can be sizable since it is $\propto g_{B-L}^4$. Therefore we can omit the reaction rates in \Eq{leptogenesis} except $\gamma_D$ and $\gamma_{Z'}$, and simplify \Eq{leptogenesis} into a single equation,
\be\label{leptogenesis_sim}
\frac{dY_{B-L}}{dz}+\left(\frac{z^4}{s_NH_N}\frac{\gamma_D}{2Y_\ell^{\rm eq}}\right)Y_{B-L}\approx\epsilon \frac{dY_N^{\rm eq}}{dz}\frac{\gamma_D}{\gamma_D+4\gamma_{Z'}},
\ee
where we have approximated $Y_N/Y_N^{\rm eq}+1\approx2$ and $dY_N/dz\approx dY_N^{\rm eq}/dz$, since $N$ is not far away from equilibrium. \Eq{leptogenesis_sim} can be solved analytically as
\begin{multline}\label{eq:kappa}
Y_{B-L}(z)\approx\epsilon\int_{z_{\rm in}}^zdz'\frac{dY_N^{\rm eq}}{dz'}\frac{\gamma_D(z')}{\gamma_D(z')+4\gamma_{Z'}(z')}\times\\\exp\left\{-\int_{z'}^zdz''\frac{z''^4}{s_NH_N}\frac{\gamma_D(z'')}{2Y_\ell^{\rm eq}}\right\},
\end{multline}
where we adopt $z_{\rm in}=1$ as the lower limit of the integral. It is very clear in above equation that $\epsilon$ and $\gamma_D$ generate the lepton asymmetry, while $\gamma_{Z'}$ tends to washout this asymmetry because it tends to push $N$ back to the equilibrium. The final generated baryon asymmetry is
\be
Y_B=\frac{28}{79}\,Y_{B-L}(z_{\rm sph}),
\ee
where $z_{\rm sph}=M_N/T_{\rm sph}$, with $T_{\rm sph}\approx130$ GeV the decoupled temperature of the EW sphaleron~\cite{Burnier:2005hp}. We have checked that \Eq{eq:kappa} matches the numerical solution of the complete equation set \Eq{leptogenesis} very well.

\begin{figure*}
\includegraphics[scale=0.25]{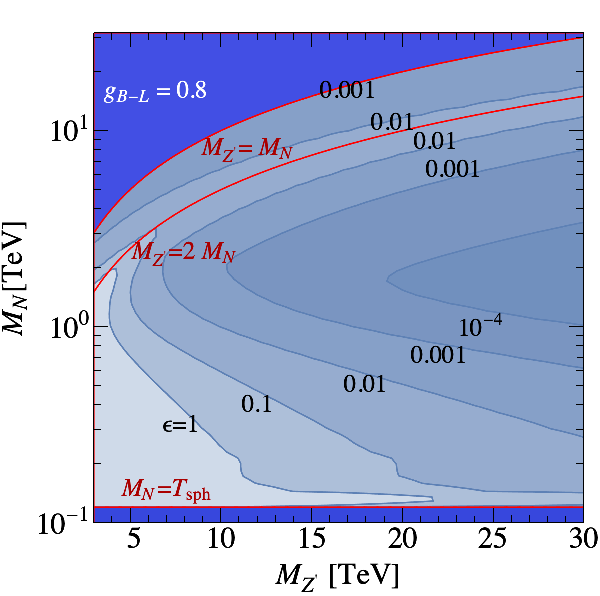}\qquad
\includegraphics[scale=0.25]{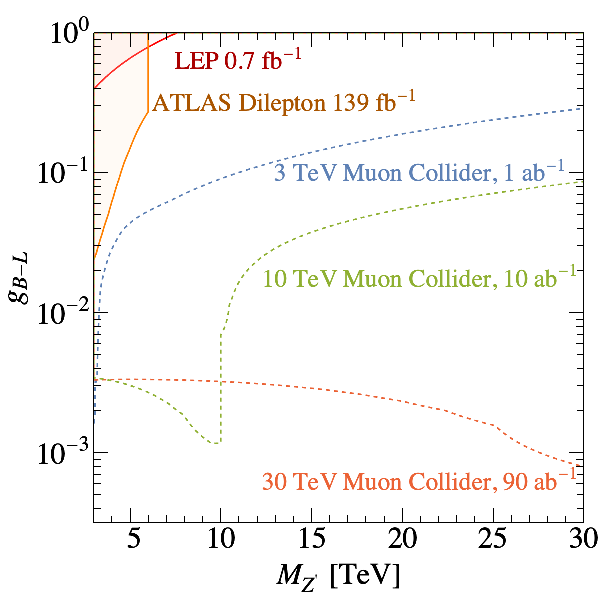}
\caption{Left: The required $CP$ asymmetry $\epsilon$ to explain the observed BAU in the mass range $T_{\rm sph.}<M_N<M_{Z'}$, for a fixed $g_{B-L}=0.8$. Only the $\epsilon$ contours for $M_{Z'}>M_N$ are shown, because for $M_{Z'}<M_N$ new scattering channels open and \Eq{eq:kappa} needs to be modified. Right: The upper limits on $g_{B-L}$ as a function of $M_{Z^\prime}$ from the LEP~\cite{Abdallah:2005ph}, ATLAS dilepton searches~\cite{ATLAS:2019erb, Chiang:2019ajm}, and projected reach from the 3 TeV muon colliders~\cite{Huang:2021nkl}, and the 10 (30) TeV muon colliders via rescaling.}
\label{fig:blz}
\end{figure*}

Given the value of $g_{B-L}$, one is able to derive the $CP$ asymmetry $\epsilon$ as a function of $(M_{Z'},M_N)$ via \Eq{eq:kappa} by the observed BAU $Y_B^{\rm obs.}\approx10^{-10}$~\cite{Ade:2015xua, Planck:2018vyg}. This is shown in the left panel of Fig.~\ref{fig:blz}, where $g_{B-L}=0.8$ is fixed. Near the line $M_{Z'}=2M_N$, the washout process $NN\to Z'\to f\bar f$ is resonantly enhanced and hence a large $\epsilon$ is needed to realize $Y_B^{\rm obs.}$. The region with $\epsilon>1$ is forbidden in the leptogenesis mechanism. We can see that there is plenty of parameter space allowed by leptogenesis for $\mO({\rm TeV})$ $Z'$ and $N$. Since this mass region is accessible at current or near future colliders~\cite{Ilten:2018crw, Cepeda:2019klc, Bagnaschi:2019djj, Deppisch:2019ldi, Deppisch:2019kvs}, some of the parameter space is already excluded by the LEP~\cite{Abdallah:2005ph} and LHC~\cite{ATLAS:2019erb, Chiang:2019ajm}, see also~\cite{Basso:2008iv,Kang:2015uoc,Cox:2017eme} searches for $Z'\to\ell^+\ell^-/jj$, as plotted in the shaded region in the right panel of Fig.~\ref{fig:blz}. Also plotted in the figure is the projected reach for $\mu^+\mu^-\to Z'^*\to\ell^+\ell^-$ and $\mu^+\mu^-\to Z'(\to\ell^+\ell^-/\nu_\ell\bar\nu_\ell)\gamma$ at the multi-TeV muon colliders taken/rescaled from Ref.~\cite{Huang:2021nkl}. As shown in the right panel of Fig.~\ref{fig:blz}, the 10~(30) TeV muon collider can reach very low $g_{B-L}$ due to the resonant enhancement where $M_{Z^{\prime}} \approx \sqrt{s}$. We are mostly interested in the parameter space that is allowed by current data but can be probed at the HL-LHC and future muon colliders; for this sake we fix $g_{B-L}=0.8$ and focus on $Z^\prime$ with $6~{\rm TeV} < M_{Z^\prime} <  30$ TeV to perform the collider phenomenology study. 

\section{Collider phenomenology}\label{sec:mc}

\subsection{Same-sign dilepton final state}

Since the SM fermions are charged under the $U(1)_{B-L}$ group, $Z'$ can be produced at the LHC via quark fusion $q\bar q\to Z'$ or at the muon colliders via $\mu^+\mu^-\to Z'\gamma/Z'Z$ (the so-called radiative return) and $\mu^+\mu^-\to Z'$. Since the collision energy $\sqrt{s}$ of a muon collider is fixed, the $Z'\gamma/Z'Z$ channel is suited for probing $M_{Z'}<\sqrt{s}$, while the $\mu^+\mu^-\to Z'$ is more appropriate to probe the off-shell region $M_{Z'}>\sqrt{s}$. As for the decay of $Z'$, we focus on the channel\footnote{The phenomenological study on resonantly produced RHNs via a heavy $Z'$ boson can also be found in Refs.~\cite{Das:2017flq,Das:2017deo,Das:2018tbd}.}
\be\label{ssdl}
Z'\to NN\to\ell^\pm\ell^\pm+{\rm jets},
\ee
channel as it is directly related to the essential ingredients of the leptogenesis mechanism: the new particles $Z'$, $N$ and the $CP$ asymmetry. By reconstructing the invariant masses of the decay products, we can find  clues for the $Z'$ and $N$ resonances; while by counting the event number difference between the $\ell^+\ell^+$ and $\ell^-\ell^-$ final states, we can probe the $CP$ violation.

\begin{figure}\centering
	\includegraphics[scale=0.25]{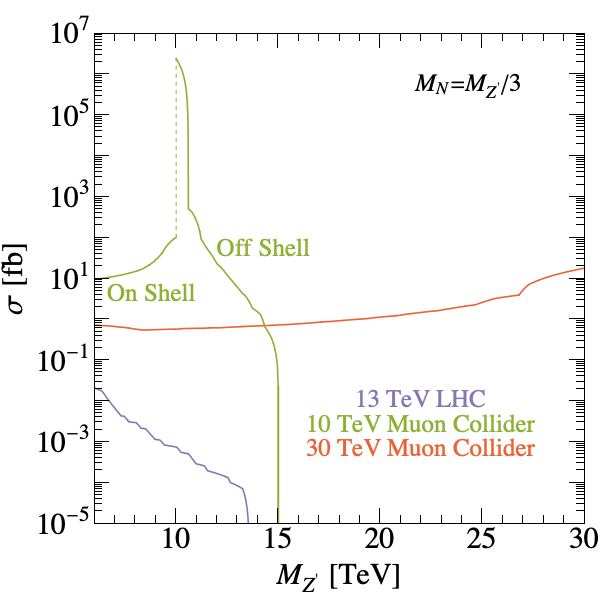}
	\caption{The production rates of $Z'\to NN$ at the 13 TeV LHC and the 10 (30) TeV muon colliders, with $M_N = M_{Z^\prime}/3$ fixed.}
	\label{fig:cszp}
\end{figure}

\begin{table*}
	\small\centering\renewcommand\arraystretch{1.2}
	\begin{tabular}{c|ccc}
		\hline\hline
		LHC  & Trigger cut [fb] &   Same-sign lepton [fb] & $W$-jet [fb]
		\\ \hline
		Signal & $\sim10^{-3}$  & $\sim$10$^{-3}$ & $\sim$10$^{-4}$   \\
		$t\bar t$ & $\sim 10^{-4}$ ($*$)  & $\lesssim10^{-7}$ &  $\lesssim10^{-10}$  \\
		$W^\pm W^\pm jj$  & $\lesssim10^{-2}$  & $\lesssim10^{-4}$ & $\lesssim10^{-7}$  \\ \hline\hline
		10 TeV muon collider  & Trigger cut [fb]   & Same-sign lepton [fb] & $W$-jet [fb]
		\\ \hline
		Signal & $\sim 1$  & $\sim 1$ & $\sim10^{-1}$  \\
		$\mu^+ \mu^- \rightarrow e^+ e^- W^+ W^-$ & $\sim10^{-2}$  & $\sim10^{-5}$ &  $\sim10^{-6}$ \\
		$\mu^+ \mu^- \rightarrow e^+ e^- W^+ W^- \gamma/Z$ & $\sim10^{-2}$  & $\sim10^{-5}$ & $\sim10^{-6}$  \\
		$\mu^+ \mu^- \rightarrow W^+ W^- j j $ & $\sim 10^{-1}$  & $\sim10^{-6}$ & $\sim10^{-9}$ \\
		%B & 10$^{-3}-$1 & 260-10000  & $M_{Z^\prime}$/3 \\
		\hline\hline
	\end{tabular}
	\caption{The cross sections of the signal and main backgrounds at the 13 TeV LHC and 10 TeV muon collider after the trigger cuts, the same-sign dilepton and $W$-jet requirements. ($*$): The decay products of $t \bar{t}$ are further required to have invariant masses $>6$ TeV. The signal process is generated at $M_{Z'}=8$ TeV, $M_N=500$ GeV, and $g_{B-L}=0.8$.}
	\label{tab:back}
\end{table*}

The cross sections of $Z'\to NN$ for various production channels are plotted in Fig.~\ref{fig:cszp} at the 13 TeV LHC, 10 and 30 TeV muon colliders where $M_{N} = M_{Z'}/3$ is fixed.\footnote{We adopt the {\tt FeynRules}~\cite{Alloul:2013bka} model file from Refs.~\cite{Amrith:2018yfb, Deppisch:2018eth}, which is also publicly available  in the {\tt FeynRules} model database, \url{https://feynrules.irmp.ucl.ac.be/wiki/B-L-SM}. The model file is then interfaced with the {\tt MadGraph5aMC@NLO-v2.8.2}~\cite{Alwall:2014hca} package for a parton-level simulation.} We can see the LHC cross section drops rapidly as $M_{Z'}$ increases, and it is only $\sigma\sim10^{-4}$ fb for $M_{Z'}=12$ TeV, yielding less than 1 event even at the HL-LHC. In contrast, the $\mu^+\mu^-\to Z'\gamma/Z'Z$ cross section increases when $M_{Z'}$ is close to the collision energy of the muon collider; for $M_{Z'}>6$ TeV, $\sigma\gtrsim10~(1)$ fb at a 10 (30) TeV muon collider, providing $\gtrsim10^5~(9\times10^4)$ events at an integrated luminosity of $10~(90)~{\rm ab}^{-1}$. The 10 TeV muon collider can probe $M_{Z'}>10$ TeV via the off-shell production of $\mu^+\mu^-\to Z'^*\to NN$. In this case, the cross section drops when $M_{Z'}$ increases, showing a tendency similar to the LHC but with much higher rates (for example $\sigma\sim10$ fb for $M_{Z'}=12$ TeV).

The RHN decays to a lepton and a vector/Higgs boson, and due to the Goldstone equivalence theorem the branching ratios are approximately
\be\begin{split}
\frac12{\rm Br}(\ell^+W^-)\approx&~{\rm Br}(\bar\nu_\ell Z)\approx{\rm Br}(\bar\nu_\ell h),\\
 \frac12{\rm Br}(\ell^-W^+)\approx&~{\rm Br}(\nu_\ell Z)\approx{\rm Br}(\nu_\ell h).
\end{split}\ee
The difference between ${\rm Br}(\ell^+W^-)$ and ${\rm Br}(\ell^-W^+)$ connects to the $CP$ violation $\epsilon$, as defined in \Eq{epsilon}. For simplicity, in the following we assume the $N$ decays exclusively to the first generation of leptons, and focus on the $NN\to e^\pm e^\pm W^\mp W^\mp$ final state, which is clean due to its same-sign dilepton feature. As we are considering $M_{Z'}\gtrsim6$ TeV, the $W^\pm$ bosons from cascade decay are typically boosted, and hence can be treated as boosted fat jets.

To trigger the signal events, we require the final state to have exactly two electrons within
\bea
&&p_T^e>100~{\rm GeV},\quad |\eta_e|<2.5,\quad {\rm (for~LHC)}\\
&&p_T^e>30~{\rm GeV},\quad |\eta_e|<2.43;\quad {\rm (for~muon~colliders)}\nn
\eea
and two jets within
\bea
&&p_T^W>500~{\rm GeV},\quad |\eta_W|<2,\quad {\rm (for~LHC)}\\
&&p_T^W>500~{\rm GeV},\quad |\eta_W|<2.43.\quad {\rm (for~muon~colliders)}\nn
\eea
The electrons are further required to be same-sign, and the jets are required to be $W$-tagged. At the LHC, the main backgrounds come from the dilepton decay of $t\bar t$ (with one lepton's charge misidentified and two QCD jets mistagged as $W$) and $W^\pm W^\pm jj$ (with jets mistagged)~\cite{Blanchet:2009bu}. While at the muon colliders, the $t\bar t$ background is subdominant; instead, the main backgrounds are $\mu^+ \mu^- \rightarrow e^+ e^- W^+ W^-$ and $\mu^+ \mu^- \rightarrow e^+ e^- W^+ W^- \gamma/Z$ (with lepton charge misidentified), $\mu^+ \mu^- \rightarrow W^+ W^- j j $ (with jets mistagged). The backgrounds from charge misidentification are significantly suppressed by the misidentification rate, which we adopt as 0.1\% based on the LHC detector performance~\cite{ATLAS:2019jvq}. Note that this is a conservative estimate as the charge identification efficiency is expected to be improved at the muon colliders.

In our parton-level study, the $W$-jet is simulated by a parton-level $W$ boson multiplying by the hadronic decay branching ratio 67.4\%~\cite{ParticleDataGroup:2020ssz} and the tagging efficiency 60\%. The backgrounds from mistagged QCD jets are then suppressed by the mistag rate, which we adopt as 5\%. The $t\bar t$ background is further suppressed by requiring the final state objects' invariant mass to be above 6 TeV, while the same cut almost does not affect the signal rate. The background rates are listed in Table~\ref{tab:back}, where the signal cross sections for $M_{Z'}=8$ TeV, $M_N=500$ GeV, and $g_{B-L}=0.8$ are also given. We can see that at the LHC the backgrounds are under control, while at the muon colliders the backgrounds are negligible after the selection cuts.

\begin{figure*}
	\includegraphics[scale=0.23]{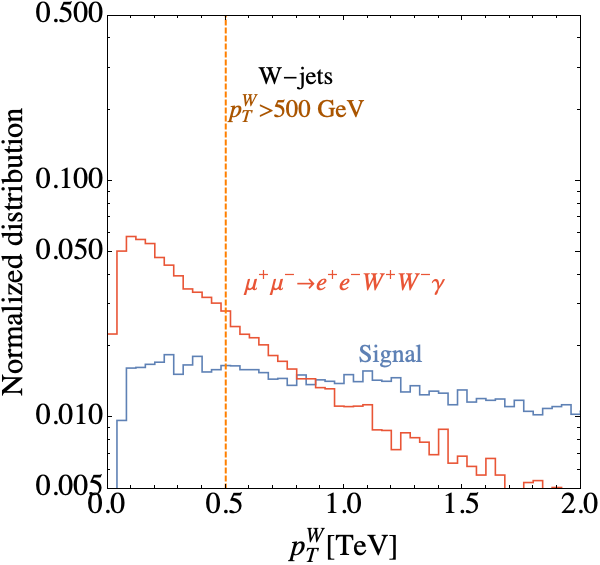}\qquad
	\includegraphics[scale=0.23]{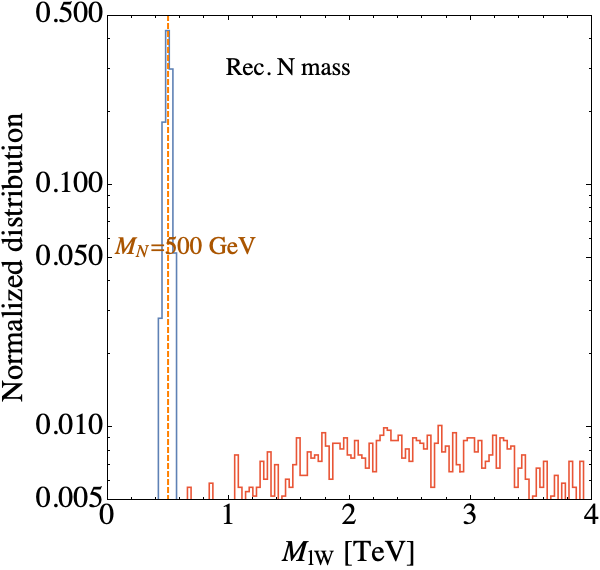}\qquad
	\includegraphics[scale=0.233]{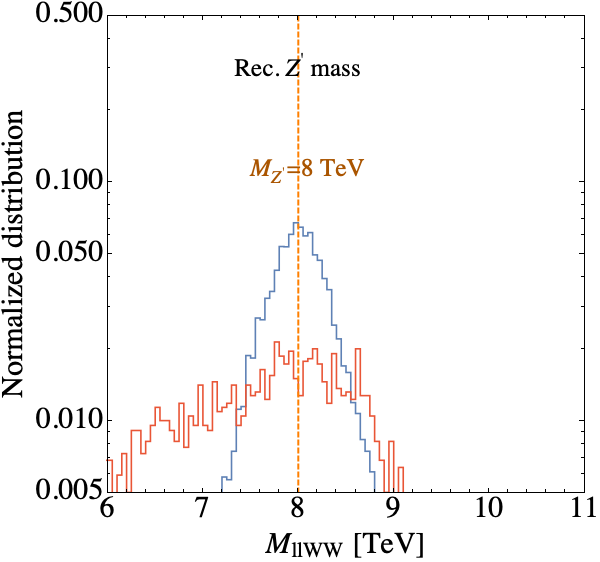}
	\caption{$p^W_T$ and the invariant mass distribution of the signal and $\mu^+ \mu^- \rightarrow e^+ e^- W^+ W^- \gamma$ at the 10 TeV muon collider for $(M_{Z^\prime} =$ 8 TeV, $M_N = $ 500 GeV).  }
	\label{fig:inv}
\end{figure*}

At the muon colliders, we can further remove the backgrounds also by putting cuts on the invariant mass of the final state particles. This is shown in Fig.~\ref{fig:inv}, where $p^W_T$ and the invariant masses of the signal and $\mu^+ \mu^- \rightarrow e^+ e^- W^+ W^- \gamma$ at the 10 TeV muon collider are illustrated for one benchmark point with $M_{Z^\prime} =$ 8 TeV and $M_N = $ 500 GeV. The jet four-momentum is smeared according to a jet energy resolution of $\Delta E/E = 10 \%$. For the reconstruction of $N$, we pair the two leptons and $W$-jets by minimizing $\chi_j^2=(M_{l_1^\pm W_1} - M_N)^2+(M_{l_2^\pm W_2} - M_N)^2$. From the figure, it is clear to see that the signal and background have very different kinematical distributions especially when reconstructing the $N$; a cut on the $\ell^\pm W$ invariant mass can significantly remove the backgrounds, even in case of $M_{Z'}>\sqrt{s}$.

\subsection{Testing Leptogenesis}

\begin{figure}\centering
	\includegraphics[scale=0.25]{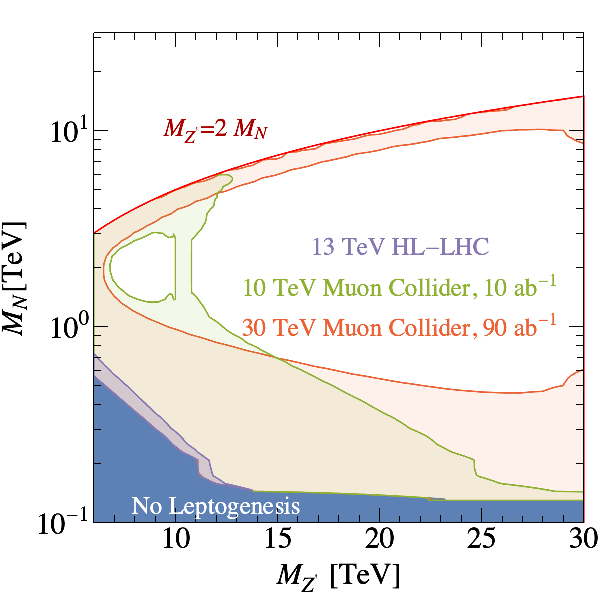}
	\caption{The regions where $\epsilon>\epsilon_{\rm min}$ that the resonant leptogenesis can be tested at the HL-LHC and the 10~(30) TeV muon collider with an integrated luminosity of 10~(90) ab$^{-1}$. $g_{B-L} =0.8$ is fixed. The dark blue region labeled as ``no leptogenesis'' is for $\epsilon>1$ or $M_N<T_{\rm sph}$.}
	\label{fig:asy_zp}
\end{figure}

We have shown in the previous subsection that the backgrounds for the same-sign dilepton channel are negligible after imposing a set of selection cuts, which however keep the signal rates almost unchanged. There are also other channels available for the $Z'\to NN$ process (e.g. $N\to \nu h/\nu Z$), which can further increase the signal significance of the model. Based on this consideration, we make the zero background assumption in this subsection to derive the projected sensitivities for the thermal leptogenesis parameter space of the $B-L$ model. The $CP$ asymmetry $\epsilon$ defined in \Eq{epsilon} is related to the asymmetry between the positive and negative same-sign dileptons as~\cite{Blanchet:2009bu}
\begin{align}
\epsilon=\frac{1}{2}\abs{\frac{N_+-N_-}{N_++N_-}},
\end{align}
where $N_\pm$ denotes the event number of the $e^\pm e^\pm W^\mp W^\mp$ final state. If we observe no asymmetry of the same-sign dileptons at colliders, upper limits can be put on the $CP$ asymmetry. This is done by assuming the number of the signal events to follow a Poisson distribution, and hence at 1$\sigma$ confidence level~\cite{ParticleDataGroup:2020ssz}
\begin{align}
N_+ =  \ave{N_+} \pm \Delta N_+=\ave{N_+} \pm \sqrt{ \ave{N_+}},
\end{align}
yielding a sensitivities of
\be\label{limit}
\epsilon_{\rm min}=\frac{1}{2 \sqrt{\ave{N_+}}}.
\ee
The $CP$ asymmetry is detectable if $\epsilon>\epsilon_{\rm min}$.

With \Eq{limit} in hand, one is able to test resonant leptogenesis at a specific collider environment. Given a set of $(M_{Z'},M_N,g_{B-L})$, the $\epsilon$ required by leptogenesis is derived by \Eq{eq:kappa}. On the other hand, the corresponding $N_+$ and hence $\epsilon_{\rm min}$ is also available by collider simulation of the $Z'\to NN$ process. If $\epsilon>\epsilon_{\rm min}$, then this parameter setup is detectable at the collider. The projected detectable parameter space with a fixed $g_{B-L}=0.8$ is plotted in Fig.~\ref{fig:asy_zp} for the HL-LHC (13 TeV, 3 ab$^{-1}$) and two setups of muon colliders (10 TeV, 10 ab$^{-1}$ and 30 TeV, 90 ab$^{-1}$) with different colors. As the signal significance is proportional to the $CP$ asymmetry $\epsilon$, the reachable regions have similar shapes to the $\epsilon$ contours in the left panel of Fig.~\ref{fig:blz}, except that there is a vertical band-like region in the 10 TeV muon collider case near $M_{Z'}\approx10$ TeV due to the enhancement from resonant production of $Z'$. The region shaded in dark blue is for $\epsilon>1$ or $M_N<T_{\rm sph}$ that the leptogenesis is not feasible. As we can see, the HL-LHC is able to probe leptogenesis for $M_{Z'}\lesssim12$ TeV with a RHN mass $M_N\lesssim600$ GeV, corresponding to the $CP$ asymmetry $\epsilon\sim0.2$. While for the muon colliders, due to the higher production rates and integrated luminosities, smaller $\epsilon\sim10^{-3}$ can be probed and a much higher mass region can be reached. At the 10 TeV muon collider, if the RHN is as light as a few hundreds of GeV, leptogenesis with $M_{Z'}$ up to 28 TeV can be probed via the off-shell process $\mu^+\mu^-\to Z'^*\to NN$. The 30 TeV muon collider can cover a wider parameter space via the on-shell $Z'Z$ or $Z'\gamma$ production for $M_{Z'}$ up to 30 TeV, and via the off-shell production for $M_{Z'}$ up to 100 TeV. We also notice that the parameter space for $M_{Z'}\sim\mathcal{O}(10~{\rm TeV})$ and $M_N\sim2$ TeV is not reachable for all collider setups, because the $\epsilon$ needed for BAU is too small (see the contours in Fig.~\ref{fig:blz}). Although larger $\epsilon$ can be generated from tuning the $\Delta M_N$ which is a free parameter, the resulting BAU will be greater than the observation value. However, a band in this region can be filled up by the 10 TeV muon collider for $M_{Z'}\sim10$ TeV, due to the resonant enhancement of the $Z'Z/Z'\gamma$ process. In summary, a considerable fraction of parameter space can be probed at the future muon colliders.

\section{Conclusion}\label{sec:con}

Both the BAU and tiny neutrino mass can be explained simultaneously within the framework of the thermal leptogenesis via the RHNs. If two of the RHNs are nearly degenerate, the scale of leptogenesis can be as low as $\sim$ 100 GeV and hence accessible at current and future colliders. In this article, we take the $B-L$ model as an example to perform a study on the collider probe of the leptogenesis mechanism. In this model, the new gauge boson $Z'$ mediates a scattering process that tends to washout the BAU. Therefore, if both $Z'$ and $N$ are at $\mathcal{O}({\rm TeV})$ scale, a sizable $CP$ asymmetry $\epsilon$ is needed to explain the observed BAU. In other words, successful (resonant) leptogenesis can be realized via TeV scale $Z'$, $N$ and a sizable $\epsilon$, which are testable at the colliders.

We choose the channel $Z'\to NN\to \ell^\pm\ell^\pm+{\rm jets}$ to probe leptogenesis, as this process involves the three key ingredients of the scenario: a new gauge boson $Z'$, the RHN $N$, and the $CP$ violation $\epsilon$ from the asymmetry between $N(\ell^+\ell^+)$ and $N(\ell^-\ell^-)$. Three collider setups are considered: 13 TeV LHC with 3 ab$^{-1}$ and 10 (30) TeV muon colliders with 10 (90) ab$^{-1}$. At the LHC, the $Z'$ is produced via the Drell-Yan process; while at the muon colliders, $Z'$ can be produced in association with a $Z'$ or $\gamma$, or via the off-shell $s$-channel fusion. We have shown that the backgrounds are safely negligible after a set of selection cuts, and hence the sensitivities for $\epsilon$ can be derived under the zero background assumption. Our quantitative study shows that leptogenesis in the $B-L$ model can be probed at the LHC for $6~{\rm TeV}\lesssim M_{Z'}\lesssim12$ TeV and $M_N\sim{\rm TeV}$. At the muon colliders, due to the higher signal rates, the projected probe limits can cover $M_{Z'}$ up to 30 TeV or even higher. For the 30 TeV muon collider, the leptogenesis can be probed up to $M_{Z^\prime} \sim 100$ TeV via the off-shell production of $Z'$. Our work demonstrates that the muon colliders can serve as a machine to efficiently probe the early Universe dynamics.

\section*{Acknowledgements}
We thank Frank Deppisch and Suchita Kulkarni for useful early discussions. We are grateful to Steve Blanchet, Zackaria Chacko and Rabindra Mohapatra for very helpful discussions. KPX is supported by the University of Nebraska-Lincoln.  WL is supported by the 2021 Jiangsu Shuangchuang (Mass Innovation and Entrepreneurship) Talent Program (JSSCBS20210213). 

\bibliographystyle{apsrev}
\bibliography{references}
\end{document}